\documentclass[aps, twocolumn,superscriptaddress, 
amsmath,amssymb,reprint,numbers,noeprint,
]{revtex4-1}

\usepackage[utf8]{inputenc}
\usepackage[english]{babel}
\usepackage{graphicx}% Include figure files
\usepackage{dcolumn}% Align table columns on decimal point
\usepackage{bm}% bold math
\usepackage{amsmath}
\usepackage{color}
\usepackage{comment}
\usepackage{array}
\usepackage{chemmacros}
\usepackage{textgreek}
\usepackage{siunitx}
\usepackage{comment}
\newcolumntype{P}[1]{>{\centering\arraybackslash}p{#1}}
\usepackage{soul}
\usepackage{xcolor}

\newcommand{\PreserveBackslash}[1]{\let\temp=\\#1\let\\=\temp}
\newcolumntype{C}[1]{>{\PreserveBackslash\centering}p{#1}}

\begin{document}

%\preprint{APS/123-QED}

%\title{Spectrally stable near-surface implanted nitrogen vacancy centers in diamond and the role of surface proximity}% Force line breaks with \\
\title{Impact of surface and laser-induced noise on the spectral stability of implanted nitrogen-vacancy centers in diamond}

\author{Srivatsa Chakravarthi}
\affiliation{University of Washington, Electrical and Computer Engineering Department, Seattle, WA, 98105, USA}%
\author{Christian Pederson}
\affiliation{University of Washington, Physics Department, Seattle, WA, 98105, USA}%
\author{Zeeshawn Kazi}%
\affiliation{University of Washington, Physics Department, Seattle, WA, 98105, USA}%
\author{Andrew Ivanov}%
\affiliation{University of Washington, Physics Department, Seattle, WA, 98105, USA}%
\author{Kai-Mei C. Fu}%
\affiliation{University of Washington, Electrical and Computer Engineering Department, Seattle, WA, 98105, USA}%
\affiliation{University of Washington, Physics Department, Seattle, WA, 98105, USA}%

\date{\today}% It is always \today, today,

\begin{abstract}

Scalable realizations of quantum network technologies utilizing the nitrogen vacancy center in diamond require creation of optically coherent NV centers in close proximity to a surface for coupling to optical structures. We create single NV centers by $^{15}$N ion implantation and high-temperature vacuum annealing. Origin of the NV centers is established by optically detected magnetic resonance spectroscopy for nitrogen isotope identification. Near lifetime-limited optical linewidths ($<$\,60\,MHz) are observed for the majority of the normal-implant (7$^\circ$, $\approx$\,100\,nm deep) $^{15}$NV centers. Long-term stability of the NV$^-$ charge state and emission frequency is demonstrated. The effect of NV-surface interaction is investigated by varying the implantation angle for a fixed ion-energy, and thus lattice damage profile. In contrast to the normal implant condition, NVs from an oblique-implant (85$^\circ$, $\approx$\,20\,nm deep) exhibit substantially reduced optical coherence. Our results imply that the surface is a larger source of perturbation than implantation damage for shallow implanted NVs. This work supports the viability of ion implantation for formation of optically stable NV centers. However, careful surface preparation will be necessary for scalable defect engineering.

%\begin{description}
%\end{description}
\end{abstract}

%\keywords{Suggested keywords}%Use showkeys class option if keyword display desired
\maketitle

%\tableofcontents

\section{Introduction}
Nitrogen-vacancy (NV) point defects in diamond combine optical addressability~\cite{doherty_nitrogen-vacancy_2013,jelezko_single_2006, robledo2011high} with long spin coherence times~\cite{abobeih_one-second_2018}, making them promising candidates for quantum networking~\cite{kimble2008quantum,wehner2018quantum}. NV centers in diamond have been used to demonstrate essential ingredients for quantum networks in recent experiments,including on-demand remote entanglement generation~\cite{humphreys_deterministic_2018,pompili2021realization}, coherent control of multiple nearby nuclear spin memories\cite{bradley_ten-qubit_2019} and memory-enhanced quantum communication\cite{kalb2017entanglement,pompili2021realization}. For networking schemes, optical coherence and photon collection efficiency are key figures of merit. Nanophotonic integration of NV centers has demonstrated potential for high collection efficiency and scalable integration~\cite{schroder_quantum_2016,wan2020large,gould_efficient_2016,chakravarthi_inverse-designed_2020} and thus should enable the scaling of quantum entanglement networks. The small mode volume needed for significant photonic coupling requires localization of NV centers to within tens of nanometers from diamond surfaces. Hybrid materials platforms~\cite{gould_efficient_2016,schmidgall_frequency_2018}, which minimize diamond fabrication, utilize evanescent coupling require NV centers in even closer surface proximity. Nitrogen ion (N$^+$) implantation followed by high-temperature annealing is a commonly utilized process for targeted spatial localization of NV centers for device integration. However, recent published results by van Dam {\it et al.}~\cite{van_dam_optical_2019} and Kasperczyk{\it et al.}~\cite{kasperczyk2020statistically} determined that centers with high optical coherence created by N$^+$ implantation and annealing are predominantly formed by implantation-induced vacancies diffusing and combining with native nitrogen. As vacancies are relatively mobile at annealing temperatures~\cite{santori_vertical_2009,davies_vacancy-related_1992,breuer_ab_1995,hu_diffusion_2002}, this result implies loss of localization and precludes deterministic photonic device integration.

The optical coherence of shallow NV centers can suffer degradation from two sources; (1) charge traps formed in the bulk from the implantation and annealing process and (2) charge traps associated with the surface or sub-surface of diamond. Ionization of charge traps produces a dynamically changing electric field which couples to the different dipole moments of the ground and excited states of the NV centers\cite{acosta_dynamic_2012, chu_coherent_2014,schmidgall_frequency_2018, ruf_optically_2019}. This effect manifests as linewidth broadening and spectral diffusion of the NV optical transitions. Since the prescription for each possible source is quite different, it is important to identify the relevant culprit. Here we show that for $\approx$\,100\,nm implant depth, it is possible to create $^{15}$NV centers with typical optical transition linewidths $<$60\,MHz. Additionally, the long-term spectral stability of the NV transitions to within 200\,MHz is demonstrated. For this implant condition, given the average NV-surface distance, we expect bulk sources to dominate optical decoherence. The observed spectral stability implies bulk sources can be overcome. 

Encouraged by the 100\,nm implantation results, we explore the possibility of implanting coherent centers closer to the surface. Shallower centers allow for enhanced optical coupling~\cite{gould_efficient_2016} for hybrid materials devices. We create NV centers at $\approx$\,20\,nm by changing the angle of implantation as opposed to varying the energy of implantation. Hence, the local damage profile around an NV center is similar to the 100\,nm implantation condition, merely rotated relative to the surface. Here we find that the optical linewidths are orders of magnitude larger and are accompanied by decreased spectral stability. 

Combined, these observations strongly imply that the proximity to the diamond surface is the dominant source of optical decoherence, and that the bulk implantation damage profile is not the limiting factor for shallow implanted NV centers.

\section{Experiment}

\subsection{Samples}
\begin{figure}[t]
\centering
\includegraphics[width=0.48\textwidth]{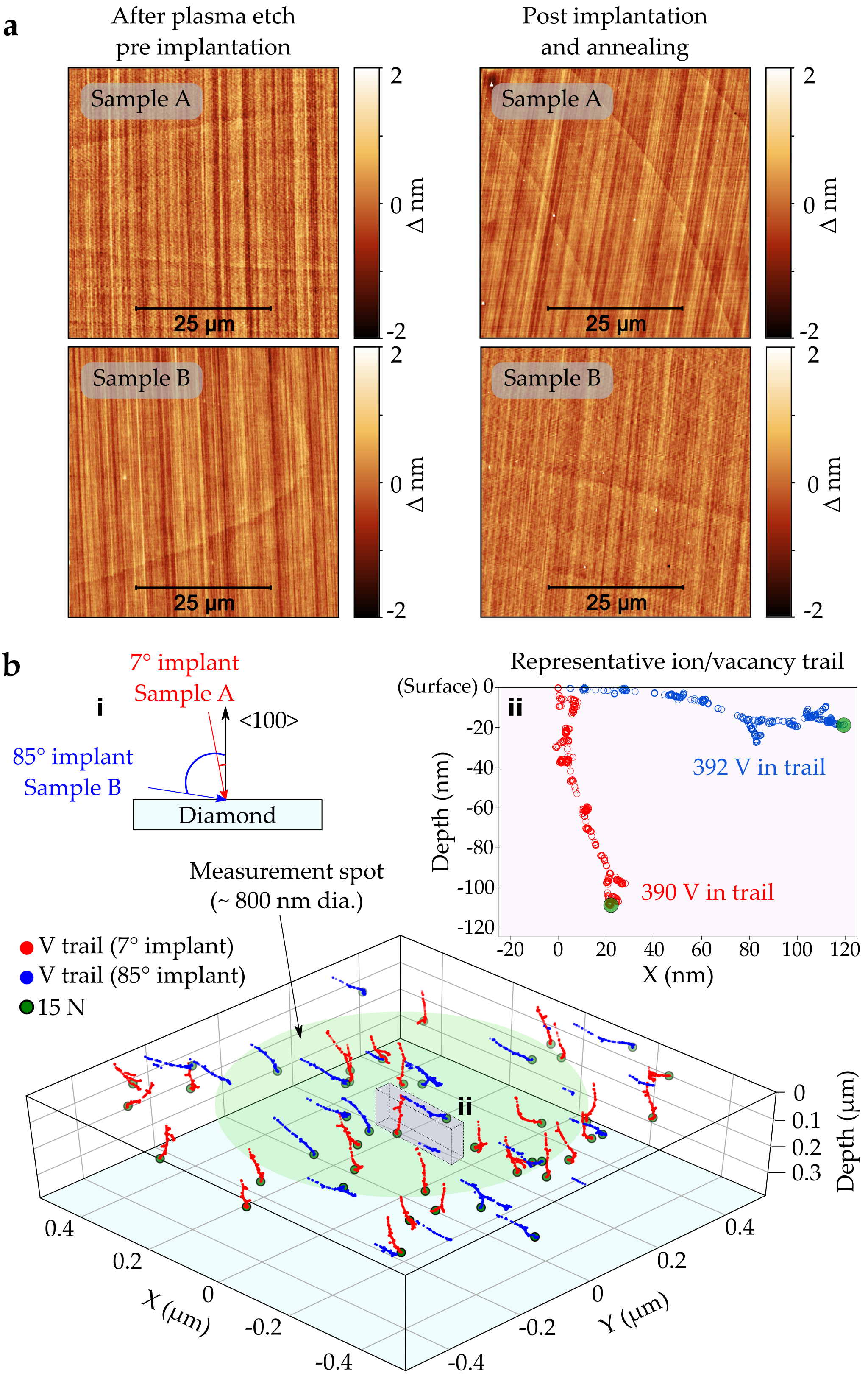}
\caption{Diamond surface preparation and simulations of implantation conditions: \textbf{a.} Morphology of the diamond surface measured by atomic force microscopy. The RMS surface roughness for sample A (B) was measured to be 0.63 (0.43)\,nm before implantation and 0.56 (0.30)\,nm after implantation and annealing. \textbf{b.} A simulation of the implantation profile obtained by SRIM for sample A (red, 7$^\circ$) and sample B (blue, 85$^\circ$) showing the damage trails and final positions of implanted $^{15}$N atoms. The green circle represents our excitation laser spot. Note that the implantation yield is low ($<$5\%), hence within an excitation spot there is a small probability each trail results in a NV center upon annealing. \textbf{Inset i.} Illustration of the implantation geometry. \textbf{Inset ii.} A cross-section of the simulation showing the ion damage trails for the two implant angles. Total vacancies generated per ion is similar for both the implant angles. However on sample B, some $^{15}$N atoms are lost due to ion scatter out of the surface.
\vspace{-2em}
}
\label{fig:afm_srim}
\end{figure}

In our primary study, to elucidate the effect of the surface on the optical properties of implanted NV centers, we utilize two identical chemical vapor deposition diamonds samples A and B (Element Six, electronic grade, N~$<$~1\,ppb, B~$<$~1\,ppb), with $\langle$100$\rangle$ surfaces. As purchased, the diamond surfaces are polished to less than 1\,nm RMS surface roughness. Both samples are processed identically unless stated otherwise. First, we etch away $\approx$5\,\textmu m from the surface using plasma reactive-ion etching to remove polishing damage~\cite{chu_coherent_2014}. We take the following precautions to avoid micro-masking that is a common occurrence during diamond etching: At each step the diamonds are cleaned in a boiling 1:1:1 mixture of H$_2$SO$_4$, H$_2$NO$_3$ and HClO$_4$ at 260\,$^\circ$C for one hour to remove organic contaminants and graphitic carbon~\cite{chu_coherent_2014}. A sapphire carrier wafer is utilized to prevent silicon contamination of the diamond surface during the etch~\cite{lee_etching_2008}. We utilize a two step Ar/Cl plasma (physical etching via sputtering) followed by O$_2$ plasma (chemical etching via oxidation) process to remove any deposited material that may result in micro-masking (process details are provided in Appendix A). The total etch duration is 45\,min of Ar/Cl$_2$ and 20\,min of O$_2$ etching. Post processing, both diamonds have nearly identical surface morphology with sample A (B) exhibiting 0.63 (0.43)\,nm RMS roughness (Fig. \ref{fig:afm_srim}a).

We implant both samples A (B) with $^{15}$N at identical energies and effective beam dosages of 85\,keV and 3e9 ions/cm$^2$. The implantation angle for samples A and B are  7$^\circ$ and 85$^\circ$ respectively. We model the effect of the different ion incidence angles on the implantation profile using the Stopping and Range of Ions in Matter (SRIM) code~\cite{SRIM}. For sample A (B), the average depth of the $^{15}$N atoms is 100$\pm$20 (21$\pm$13)\,nm below the surface. Although the effective beam dosage is the identical for both samples, approximately 41\% of the incident ions are back-scattered for sample B. This back-scattering is a geometric consequence of rotating the damage profile relative to the surface, such that some of the scattered ions escape the diamond surface. Hence the final $^{15}$N density in sample B is predicted to have 60\% the density of sample A.

The samples are vacuum annealed at $<$\,1.4$\times$10$^{-7}$\,mbar, 1100\,$^\circ$C for two hours with long ramp times as described in Ref.~\cite{chu_coherent_2014, van_dam_optical_2019}. This is followed by a short two hour anneal at 435\,$^\circ$C under O$_2$ flow to oxygen terminate the surface and stabilize the negative charge state~\cite{fu_conversion_2010,yamano_charge_2017} of the near-surface NV centers.

In addition, we characterize four supplemental diamond NV implantation samples (C-F) to support the reproducibility of the primary study. These diamond substrates have identical specifications (Element Six, electronic grade) but are sourced from different growth runs. Pre-implantation, all samples are processed as detailed in this section. The specifics of implantation/annealing conditions for each sample are provided in the table accompanying Fig.~\ref{fig:old_ple}. 

\begin{figure*}[t]
\centering
\includegraphics[width=\textwidth]{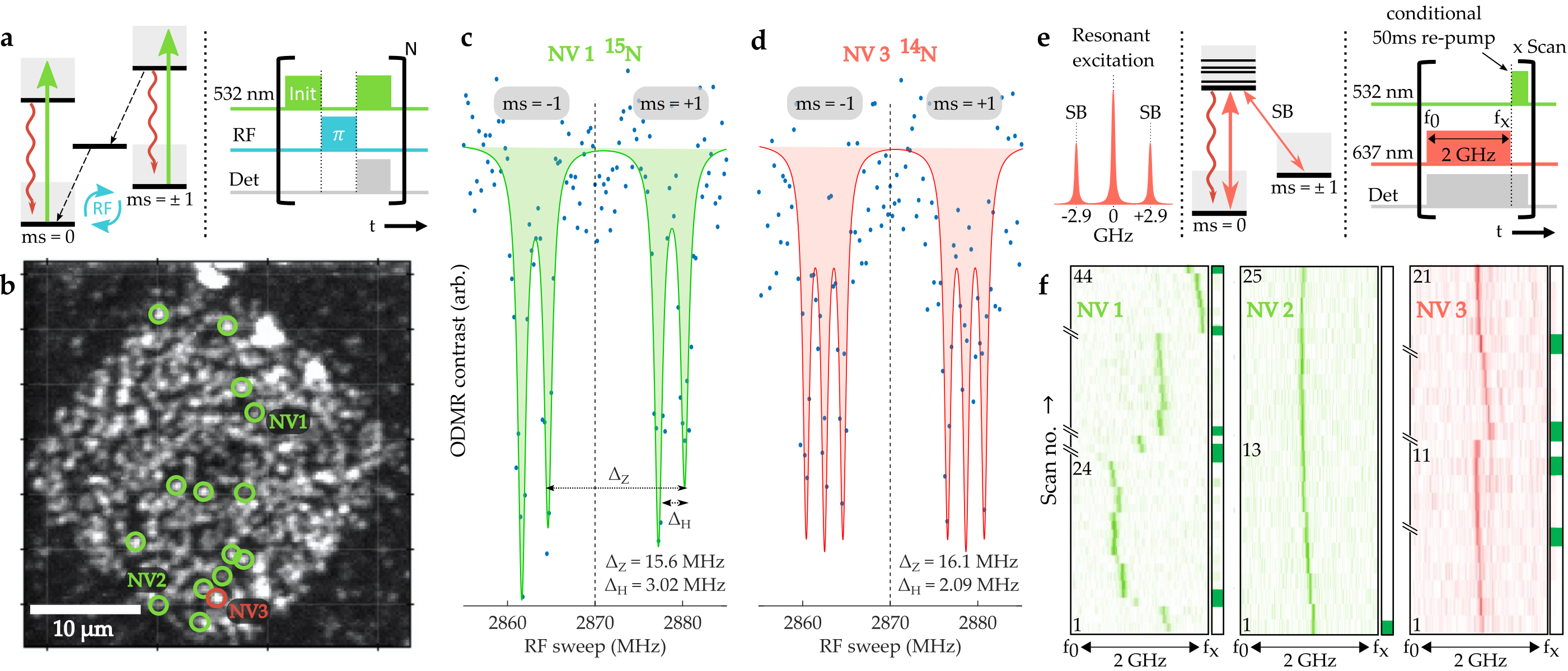}
\caption{Correlated NV ODMR (RT) and PLE (T$<$12K) measurements on sample A: \textbf{a.} Pulsed ODMR scheme utilized to identify the N isotope. Laser and radio-frequency (RF) pulses are generated by an acousto-optical-modulator and RF switch, respectively. (Laser power 0.8-1.0\,mW, spot diameter $\sim$800\,nm.) Pulses and photon collection are triggered by a programmable pulse generator.  \textbf{b.} Confocal PL map of the implanted region with the measured NV centers indicated by their isotope (green $^{15}$N, red $^{14}$N). \textbf{c.} ODMR spectra for the marked NV incorporating implanted $^{15}$N. \textbf{d.} ODMR spectra for the marked NV incorporating grown-in $^{14}$N. \textbf{e.} Resonant excitation (PLE) scheme utilized for characterizing the optical coherence of the marked NV centers. Sidebands at 2.9\,GHz are added (using an electro-optic modulator) to the scanning resonant laser to counteract optical spin pumping. Upon detection of an NV$^-$ to NV$^0$ ionization event (indicated by lack of NV$^-$ PL) a 50\,ms off-resonant green re-pump is used to reset the NV charge state. \textbf{f.} Time traces of PLE scans measured at 10.5\,K for three NV centers. Off-resonant re-pump between scans are indicated by green squares along the right column, this induces large spectral jumps in many $^{15}$NV centers (e.g. NV 1). The '$\backslash\backslash$' markers along the scan axis indicate discarded scans where no NV PL is observed. The PLE traces for $^{15}$NV centers typically show long periods of spectral stability between re-pump pulses.}
\label{fig:odmr_ple_sampleA}
\end{figure*}

\subsection{Measurements}
A confocal microscope comprising a 532~nm DPSS laser and 60X (NA=0.7) objective lens is used to scan over 80x80\,\textmu m$^2$ areas using a piezo stage. A polarizing beamsplitter with an automated half-wave plate is used in the excitation path to preferentially excite a given NV orientation. 

To identify the nitrogen isotope associated with each NV center, we use optically-detected magnetic resonance (ODMR) spectroscopy. For an NV in the ms\,=\,$\pm$1 ground spin-sublevel, reduced photoluminescence (PL) is observed upon off-resonant excitation due a small likelihood ($\approx$ 20\%) of relaxation through the dark inter-system crossing transition~\cite{goldman_state-selective_2015, tetienne_magnetic-field-dependent_2012} (dotted line, Fig.~\ref{fig:odmr_ple_sampleA}a). The samples are placed in a weak magnetic field ($\approx$ 5\,G) that splits the ms\,=\,$\pm$1 ground spin-sublevels. RF excitation is delivered via a small copper loop (radius 0.3\,mm) suspended $\approx$\,50\textmu m above the diamond sample. A short (5\,\textmu s) off-resonant 532\,nm laser pulse initializes the NV into the ms\,=\,0 spin state. Next, a radio-frequency (RF) $\pi$-pulse (0.8\,\textmu s) rotates the NV spin state before time resolved NV PL is recorded during the subsequent short (5\,\textmu s) readout laser pulse. The $\pi$-pulse area is initially calibrated by performing a Rabi experiment. The pulse sequence is repeated while sweeping the RF driving frequency over all the NV ground state spin transitions. The resulting two (three) dip PL intensity spectrum corresponds to the $^{15}$N ($^{14}$N) NV-N ground state hyperfine interaction~\cite{doherty_nitrogen-vacancy_2013,yamamoto_isotopic_2014}, indicating that the NV incorporates an implanted or grown-in nitrogen atom (Fig.~\ref{fig:odmr_ple_sampleA}c,d).  ODMR spectra are measured at room-temperature for randomly sampled NV centers in the implantation region (Fig.~\ref{fig:odmr_ple_sampleA}b) and fit to a three ($^{14}$NV) or two ($^{15}$NV) dip Lorentzian. The positions of the sampled NV centers are recorded. The samples are then cooled to $<$12\,K in a close cycle cryostat for spectral characterization of the selected NV centers.

Low-temperature NV$^-$ PL spectra under CW 532\,nm excitation provides the inhomogeneous distribution of the ZPL transition (Fig.~\ref{fig:zpl_dist}) arising from variations in the local strain and electric field environment of individual centers. The NV charge state ratio (NV$^-$/NV$^0$) is also recorded as a function of the excitation intensity for a subset of centers (Appendix C). Additionally, high resolution photoluminescence excitation (PLE) spectroscopy provides insight into the optical coherence and temporal spectral stability of individual NV centers. In PLE measurements, a narrow-band tunable laser is scanned across the NV$^-$ ZPL while collecting the NV$^-$ phonon-sideband PL (650\,nm to 800\,nm) (Fig.~\ref{fig:odmr_ple_sampleA}e). The resonant laser and accompanying 2.9\,GHz sidebands simultaneously drive the \{ms=0, ms=$\pm$1\} $\rightarrow$ \{Ex, Ey\} transitions~\cite{santori_coherent_2006,tamarat_spin-flip_2008,fu_observation_2009}. 

From the PLE spectra we collect statistics on the ZPL single-scan linewidth as well as the scan-to-scan variation in the ZPL frequency. During PLE we can sometimes observe a loss of the NV$^-$ PL signal due to ionization to the NV$^0$ state; to reinitialize into the NV$^-$ charge state we apply a short low power 532\,nm repump pulse (50\,ms) between scans (as indicated by the green markers in Fig.~\ref{fig:odmr_ple_sampleA}f). The interval between repump pulses is an additional indicator of the stability of NV$^-$ charge state.

\section{Correlated ODMR and PLE spectroscopy}

On sample A, ODMR was performed on 32 NV centers with 26 centers identified as $^{15}$NV, one as $^{14}$NV; remaining five NVs could not be conclusively identified. Similarly, on sample B, ODMR was performed on 38 NV centers with 27 centers identified as $^{15}$NV, two as $^{14}$NV; the remaining nine NVs could not be identified. The observed total NV density for sample A (B) is 1.2/\,\textmu m$^2$ (0.3/\,\textmu m$^2$) corresponds to an implantation conversion yield of 4\,\% (1\,\%). For both these samples, no grown-in NVs were observed in a 80\,\textmu m\,$\times$\,80\,\textmu m area at a depth of 50\,\textmu m implying very low native N$_s$ density~\cite{edmonds_production_2012}. Considering the natural abundance of $^{15}$N (0.4\,\%) and the $^{15}$NV/$^{14}$NV ratio $r$ for both  samples ($r{_\textrm{A}}$\,=\,26, $r_{\textrm{B}}$\,=\,13.5), it is clear that for our diamond substrates and implantation conditions that NV formation incorporating implanted nitrogen is favored.

\begin{figure}[t]
\centering
\includegraphics[width=0.48\textwidth]{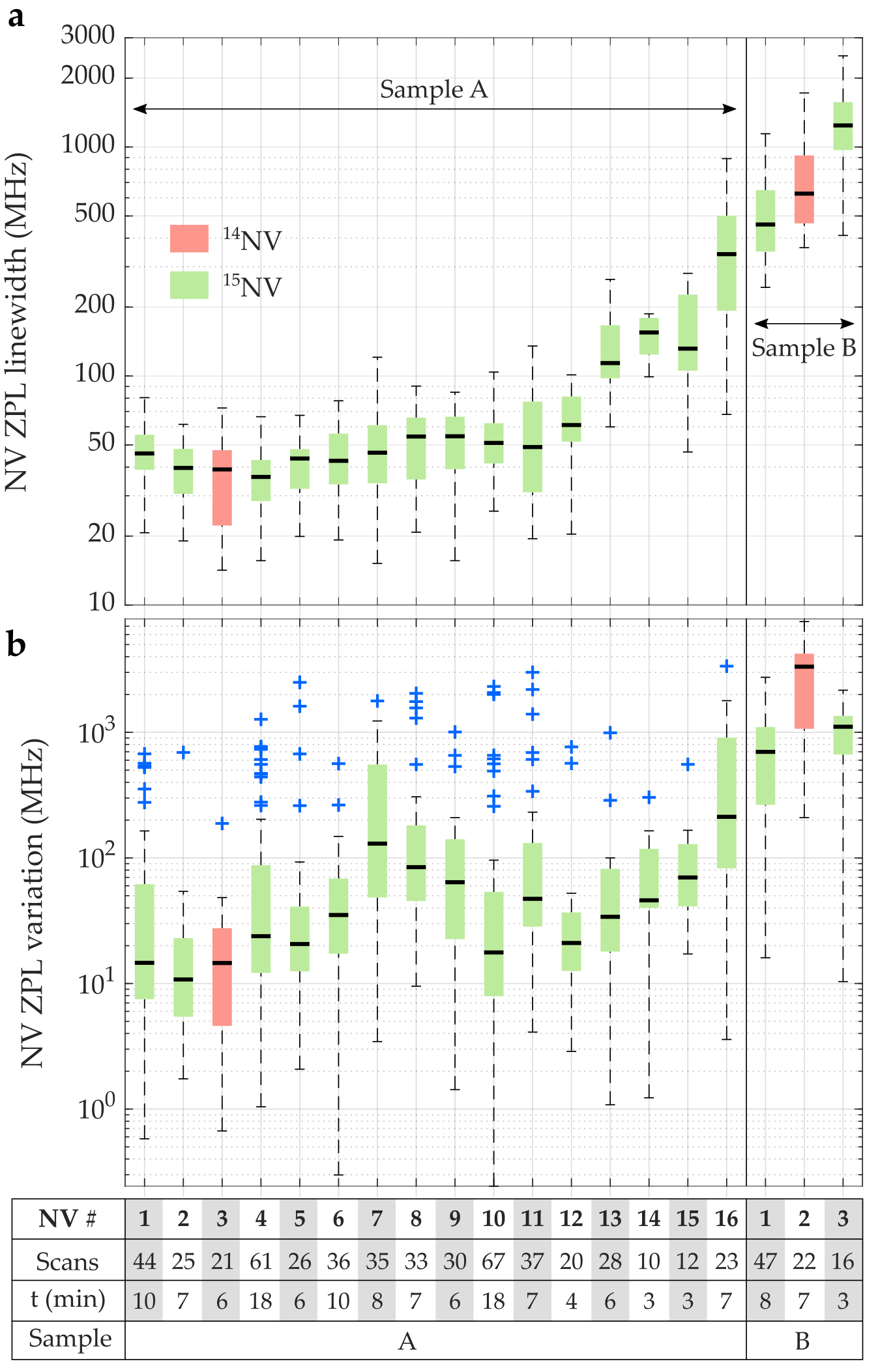}
\caption{Photoluminescence excitation characteristics of measured NV centers on samples A and B: \textbf{a.} The fitted Lorentzian FWHM distributions for each observed NV. The color box and black marker represent the interquartile range and median linewidth respectively. The total number of scans and measurement duration comprising each distribution is recorded in the accompanying table. The PLE traces for NV centers 1 to 3 are also shown in Fig.~\ref{fig:odmr_ple_sampleA}f. \textbf{b.} The distributions of the scan-to-scan change in the center frequency of the fits, representing spectral variation. Off-resonant re-pump pulses are only applied when NV ionization is detected. On average, there are six re-pump events over a measurement duration of 10\,min. The blue markers indicate large spectral jumps after re-pump pulses.}
\label{fig:ple_analysis}
\end{figure}

First let us consider sample A. The single NV low-temperature ZPL spectra under off-resonant 532\,nm excitation is typically spectrometer resolution limited ($\Delta \lambda =$ 0.021\,nm). PLE spectroscopy reveals that both $^{15}$N and $^{14}$N centers typically exhibit near lifetime-limited linewidths. A sequence of laser scans over a span of $\approx$\,10 minutes (each scan is 4 to 8\,s in duration) gives us a median linewidth of $<$\,60\,MHz for 12 out of 16 centers (Fig.\ref{fig:ple_analysis}a). This linewidth is computed by individually fitting each scan to a Lorentzian. The laser intensity is set between 30 to 60\,nW with a scan rate of 1 to 2.5\,GHz/s. A linewidth power dependence measurement was performed on two centers to ensure that the observed linewidths are not significantly power-broadened in our intensity range. The PLE results are summarized in Fig.~\ref{fig:ple_analysis}a; the solid boxes mark the fitted linewidths between first and third quartile. This interquartile range is an indicator of spectral diffusion ($\Delta\,\nu$) of the NV ZPL during a single resonant laser scan (on the timescale of ms). 

Additionally, by tracking the ZPL frequency between scans, we can characterize the long term spectral stability (on the timescale of s). We calculate this spectral variation ($\Delta\,\omega$) by recording the change in ZPL frequency between subsequent scans (Fig.\ref{fig:ple_analysis}b). Here, we emphasize that an off-resonant re-pump pulse is only applied when no NV PL is detected (i.e. NV$^-$ has ionized to NV$^0$) emulating emerging NV quantum networking protocols \cite{pompili2021realization, humphreys_deterministic_2018}. The median spectral variation is typically $<$\,100\,MHz, for long periods (60 to\,300 s) between re-pumps. We observe that most $^{15}$NV centers experience large spectral jumps ($\approx$\,500\,MHz) after an off-resonant re-pump pulse. These jumps occur in 95\% of the re-pump events. In Fig.\ref{fig:ple_analysis}b, the blue markers indicate re-pump triggered spectral jumps. This is the only metric where we see a clear advantage for the $^{14}$NV also observed on sample A (Fig.~\ref{fig:ple_analysis}, red). It is unclear at this time if the re-pump triggered perturbation originates at the surface (with the single $^{14}$NV lying deeper within the sample) or from local implantation damage. Nevertheless, it may still be mitigated with the use of a low-power resonant NV$^0$ re-pump pulse~\cite{siyushev_optically_2013}.

\begin{figure}[t]
\centering
\includegraphics[width=0.48\textwidth]{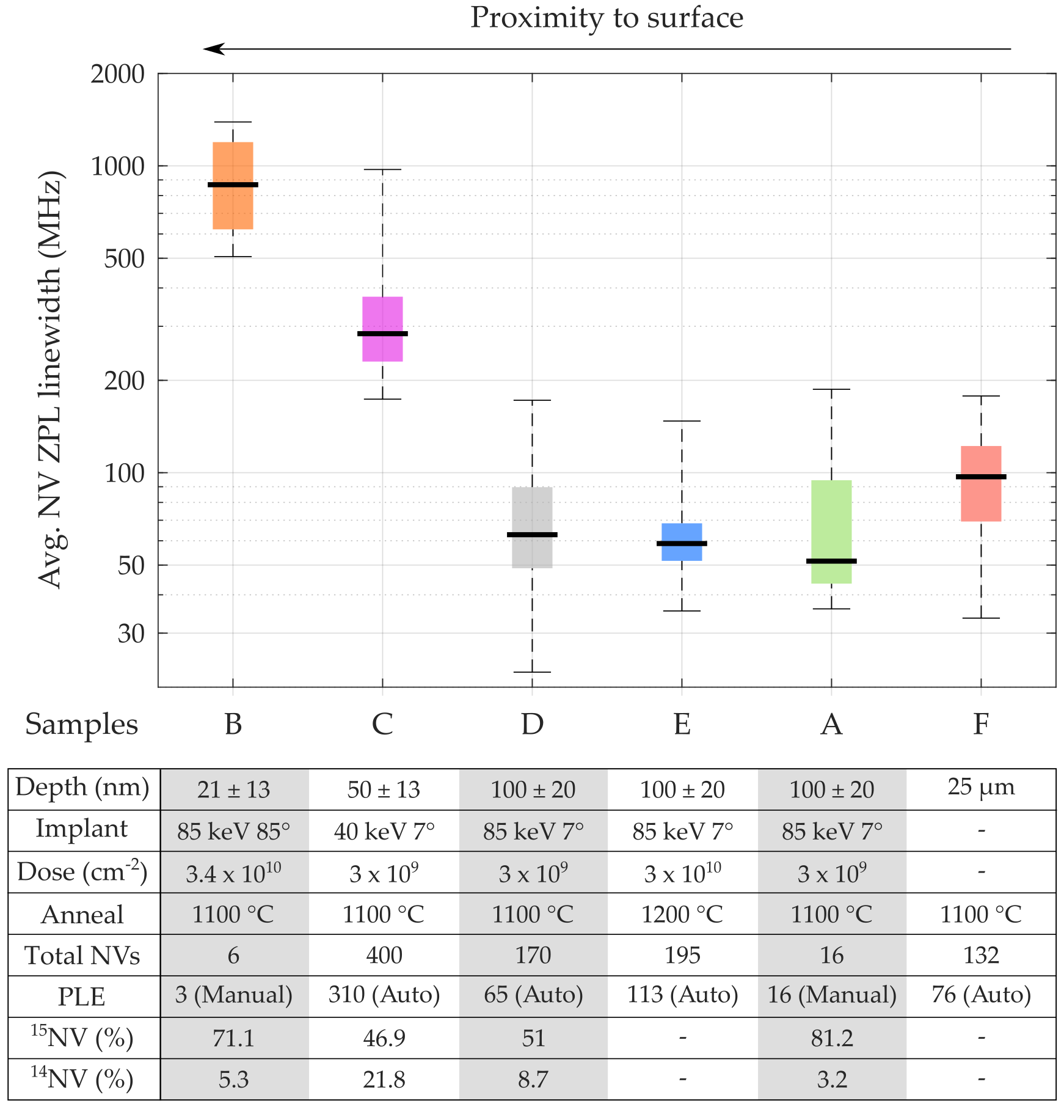}
\caption{Automated ODMR and PLE measurements are performed on four additional samples (C to F). We are not able to track the NV centers between ODMR and PLE with the automated protocol. ODMR $^{15}$NV/$^{14}$NV ratio suggests predominance of $^{15}$NVs across all implanted samples. Similar to the previous dataset (Fig.~\ref{fig:ple_analysis}), the average linewidth for each individual NV is computed as the mean FWHM obtained by fitting each of the 30 frequency scans to a Lorentzian. Automated PLE scans performed on deep grown-in NVs (25\,\textmu m from surface, sample F) are used as a reference.}
\label{fig:old_ple}
\end{figure}

Sample B reveals a different story, the low-temperature ZPL spectra under off-resonant 532\,nm excitation for individual NV centers is much broader (0.02 to 0.25\,nm;). Given that the lattice damage profile is similar to sample A, this spectral broadening could be attributed to rapid fluctuations of surface charges effectively Stark-tuning the NV centers within the exposure duration of the spectra. Such rapid ZPL fluctuations make resonant 637\,nm PLE measurements very challenging. Of the six NV centers randomly sampled, only three showed  PLE signal. All three NV centers (two $^{15}$N and one $^{14}$N) exhibit broad median linewidths (0.5\,GHz to 1.2\,GHz) and increased spectral variability (Fig.~\ref{fig:ple_analysis}).

Finally, we confirm that the NV$^-$ charge state is preferred across the full range of optical powers (15 to 600 \textmu W of 532\,nm excitation) for both samples A and B (Appendix C).

\section{Automated spectroscopy}
To corroborate the data from our primary samples (A, B), we present automated ODMR, PLE and low-temperature off-resonant PL spectroscopy datasets on four other samples (C to F). Our automation procedure allows us to sample hundreds of NV centers, however we are unable to track individual NV centers between PL, PLE and ODMR datasets. Details of the automated measurement protocol are provided in Appendix B.

\begin{figure}[t]
\centering
\includegraphics[width=0.48\textwidth]{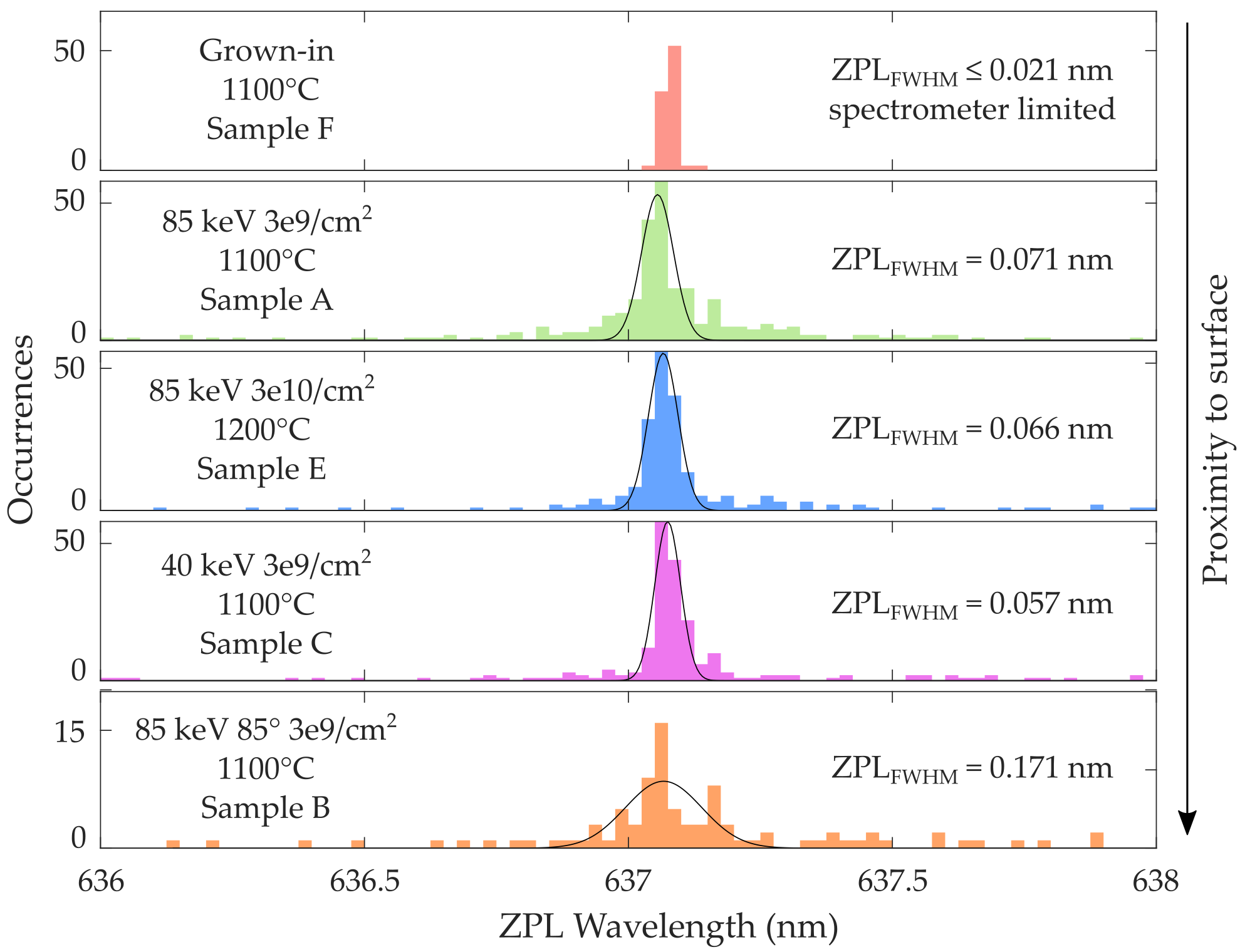}
\caption{The NV ZPL inhomogeneous distribution. The histogram is generated by recording the center wavelength of all peaks observed with 532\,nm excitation from locations within an implantation region (except the grown-in reference NV centers) on the respective samples. Note that no distinction is made of the different transitions associated with the NV excited state spin sublevels. The bin size is 0.025\,nm, and the spectrometer resolution is 0.021\,nm. The histogram data is fitted to a Gaussian to extract the full-width half max (FWHM) of the distribution. Sample B show an obvious deviation from other samples.}
\label{fig:zpl_dist}
\end{figure}

First, let us consider samples D and E with similar implant conditions ($^{15}$N, 85\,keV 7$^\circ$) to sample A. Uncorrelated ODMR measurements on sample D show most centers are $^{15}$N. The measured average ZPL linewidth distribution (Fig.~\ref{fig:old_ple}) of hundreds of NV centers, tracks well with the dataset from sample A, indicating reproducibility of narrow-linewidth $^{15}$NV centers. In the ideal case, the linewidth of NV centers created through ion implantation and annealing would be equal to the linewidth observed in background NV centers distributed throughout the sample. No background NV centers could be identified in either samples (A, B) and a low density prohibited automated measurements in samples (C, D, E). Automated PLE measurements on native NV centers 25\,\textmu m within sample F, a similar electronic-grade sample that has undergone high-temperature annealing (with no implantation), serve as our reference. From the data presented in Fig.~\ref{fig:old_ple}, the average NV linewidth distribution of all the 85\,keV, 7$^\circ$ implant samples are in agreement with the reference sample.

Next, we examine the shallow implantation samples. The average ZPL linewidth distribution in Fig.~\ref{fig:old_ple}, shows that both the shallow NV samples C (40\,keV 7$^\circ$) and B (85\,keV 85$^\circ$) exhibit decreased optical coherence. This is despite the fact that initial implantation damage for sample C is significantly lower compared to samples A, D and E (85\,keV 7$^\circ$). We can use the average number of vacancies (V) generated per implanted ion trail as an analogue for local lattice damage. From SRIM~\cite{SRIM} simulations, sample C incorporates 203 V/ion whereas samples A, B, D and E incorporate 390 V/ion. This provides further evidence that the broadening seen for the shallow implants is unrelated to the implantation damage at these energies and dosages and instead is a result of charge traps associated with the surface.

In addition, low-temperature PL spectra (off-resonant excitation) from a large number of NV centers provides the inhomogeneous ZPL distribution (Fig.~\ref{fig:zpl_dist}). Regardless of the variations in the total implant damage (V/\textmu m$^3$), the inhomogeneous NV ZPL distributions are similar for samples A, C and E (Gaussian fit FWHM\,=\,52.5\,GHz, 42.1\,GHz and 48.8\,GHz respectively). The grown-in NV centers exhibit a narrower ZPL spread (FWHM\,$\leq$\,15.4\,GHz, spectrometer resolution limited). Finally, the ZPL distribution for sample B is much broader (FWHM\,=\,126\,GHz) than any of our other samples. A comparison suggests that there is still a small amount of residual strain remaining in the implantation samples.

\section{Conclusion}
Lattice damage from the ion-implantation process introduces localized perturbations of the defect environment. Our result with the 85\,keV 7$^\circ$ implant samples indicates much of this damage can be annealed out for an $^{15}$NV created directly from an implanted nitrogen. We observe that $^{15}$NV centers exhibit long periods of spectral stability, wherein their spectral characteristics are akin to optically coherent grown-in NV centers. The implanted $^{15}$NV and grown-in centers exhibit comparable median optical linewidths. However, we do see distinct advantages for grown-in centers in terms of behavior under off-resonant green re-pump and inhomogeneous ZPL distribution. For $^{15}$NV centers, green re-pump pulses used to reinitialize the NV charge state introduce large spectral variation ($\approx$\,500\,MHz). This behavior hints at why our $^{15}$NV linewidths are qualitatively different from van Dam {\it et al.}~\cite{van_dam_optical_2019} and Kasperczyk {\it et al.}~\cite{kasperczyk2020statistically}. They utilize re-pump pulses every scan, which we can see would cause linewidth broadening into GHz. In fact, an analysis considering only the PLE scans immediately after the repump pulse indicates our results in Sample A are consistent with the 400\,nm 400\,keV $^{15}$N implanted sample in van Dam {\it et al}. Further studies are necessary to pin down the source of the re-pump triggered variation. Nevertheless, for the $^{15}$NV centers, observation of long periods of spectral stability between re-pumps shows promise for implementation of quantum networks with integrated photonic devices. 

%Utilizing the re-pump broadened linewidth metric employed by van Dam {\it et al.}~\cite{van_dam_optical_2019}, by consolidating data from Fig.~\ref{fig:ple_analysis}a and b, we obtain overall $^{15}$NV linewidths $\lesssim$\,2\,GHz. This is consistent with results from Ref.~\cite{van_dam_optical_2019,lekavicius_diamond_2019}.
%We note that in Ref.~\cite{van_dam_optical_2019,lekavicius_diamond_2019}, the NV behavior in the absence of re-pump induced spectral instability was not reported.
%This work corroborates prior work by Ref.~\cite{chu_coherent_2014} showing stable NV center emission. However in Ref.~\cite{chu_coherent_2014}, the isotope of the N and thus the NV origin, was not determined. Our results are in contrast with more recent work by Ref.~\cite{van_dam_optical_2019} with isotope identification in which stable NV emission was not observed for implanted nitrogen.

Further, we show that this optical stability rapidly degrades with increasing proximity to the diamond surface, corroborating measurements by Ref.~\cite{ruf_optically_2019} of NV centers in thin diamond membranes which showed strong correlation of reduced NV spectral stability with decreasing membrane thickness, not accompanied by a change in the NV strain environment. Ref.~\cite{ruf_optically_2019} suggests that the additional NV dephasing may be attributed the  Ar/Cl$_2$ plasma etch process even for micron-scale thick samples. Our past work has seen a similar effect on implanted centers~\cite{chakravarthi_inverse-designed_2020}. Indeed, Ref.~\cite{lekavicius_diamond_2019} show that $\approx$\,1\,\textmu m thin diamond membranes with improved diamond surface quality (etched by a soft graded O$_2$ plasma) can host NV centers with narrow linewidths.

Our conclusions fit into a larger narrative regarding the source of degradation of other properties near surfaces as well, such as NV $T_2$. Recent work by Ref.~\cite{sangtawesin_origins_2019} correlating the reduction of NV $T_{2}$ within 20 nm of the surface has shown that the $T_{2}$ for NV center within 10 nm of the surface can be enhanced an order of magnitude by preparing diamonds with smoother surfaces and well-ordered oxygen termination. Further work would need to be done to determine whether the spin-bath responsible for $T_{2}$ degradation is related to the charge traps we infer in our optical measurements, however both suggest that solving the surface interaction problem is more important than fixing residual implantation damage.

\begin{acknowledgments}
We would like to thank N.P. de Leon and R. Hanson for helpful discussions; undergraduate REU students K. Crane and M. Chamberlain for assistance with measurement automation; E.R. Schmidgall for help with fabrication process development and pre-implantation etching of our diamond samples. This material is based upon work supported by the National Science Foundation under ECCS-1807566. The diamond samples were processed at the Washington Nanofabrication Facility, a National Nanotechnology Coordinated Infrastructure (NNCI) site at the University of Washington which is supported in part by funds from the National Science Foundation award NNCI-1542101.
\end{acknowledgments}

\appendix{
\section{Plasma processing details}
We utilize an Oxford Plasmalab-100/ICP-180 etcher. The samples are sandwiched between two sapphire slides held in place with a drop of Crystalbond-509 (applied as a solution in acetone) on a 100\,mm sapphire carrier wafer. The total etch duration is broken up into multiple etch cycles involving 5\,min of plasma processing followed by a 3\,min no plasma cooldown phase. This ensures the diamond sample is maintained near the processing temperature. The samples remain in the etcher through the entire two step process. The etch parameters are provided in table.~\ref{table:plasma}.

\section{Automated spectroscopy procedure}
Confocal scans with off-resonant 532\,nm excitation are utilized to generate an NV PL intensity map of the region of interest. First, individual NV centers are identified by image processing (peak prominence detection) and their x, y and z piezo positions are registered. The linearly polarized excitation is optimized for one set of NV orientations. ODMR, PLE and low-temperature off-resonant spectra datasets are acquired by iterating through the registered NV centers. During the iteration process, at each registered center the piezo x, y and z positioners are cycled through three PL optimization sweeps to correct for microscope drift. Before proceeding with pulsed ODMR at RT, a Rabi experiment is manually performed to extract the RF pi-pulse specifications for the dataset. This is followed by ODMR performed at RT.  

\renewcommand{\arraystretch}{1.4}
\begin{table}[t]
\begin{tabular}{|c|C{0.1\textwidth}|C{0.1\textwidth}|}
\hline
Parameter                       & Ar/Cl$_2$    & O$_2$      \\ \hline
\hline
RF power (W)                    & 240       & 50            \\ \hline
ICP power (W)                   & 320       & 1500          \\ \hline
DC bias (V)                     & 530       & 150           \\ \hline
Chamber Pressure (mTorr)        & 9         & 25            \\ \hline
Ar flow (sccm)                  & 32        & 0             \\ \hline
Cl$_2$ flow (sccm)              & 20        & 0             \\ \hline
O$_2$ flow (sccm)               & 0         & 20            \\ \hline
Chuck temperature ($^\circ$C)   & 15        & 15            \\ \hline
Total duration (min)            & 45        & 20            \\ \hline
\end{tabular}
\caption{Plasma processing parameters for pre-implantation etch of all diamond samples.}
\label{table:plasma}
\end{table}
\renewcommand{\arraystretch}{1}

Next, the samples are cooled in a closed cycle 10\,K cryostat (Janis CCS-XG-M/204N) for the PLE dataset. Because we switch microscopes between automated ODMR (RT) and PLE (LT), the datasets are not correlated. Hence for our primary correlated dataset (samples A, B) consecutive ODMR and PLE measurements were performed manually on the same microscope. For each center, PLE is performed in two steps, coarse and fine scans. First, the coarse scans utilize the full range of a New Focus velocity tunable laser ($\approx$\,85\,GHz) to identify the ZPL frequency. Then a set of 30 scans are performed across the identified ZPL (scan range = 5\,GHz) with high resolution ($\Delta\,\nu$=10\,MHz). For samples C, D and F, a 50\,ms off-resonant re-pump is applied at the end of each scan. For sample E, re-pump pulse is only applied if no NV PL is observed during scan (i.e. indicating NV has ionized to the neutral charge state). In post-processing, each scan is fitted to a Lorentzian, and the average FWHM of the fits is calculated. During analysis, a set of preset criteria (peak intensity, fitted FWHM, fit shape) are used to discard scans with ionization events. The average FWHM distribution thus computed for all NV centers in the dataset is shown in Fig.~\ref{fig:old_ple}. No sideband or microwave driving was used for the automated PLE datasets.

Finally, off-resonant ZPL spectra is collected for registered NV centers with a 1800g Princeton Acton 2750 spectrometer ($\Delta\,\lambda$=0.0208\,nm). Here, if multiple peaks are observed in individual spectra, they are recorded as independent peaks. We do not distinguish the different transitions associated with the NV excited state spin sublevels. To identify the excited state structure with confidence would require a confirmation of a single NV in the excitation spot, obtained via a photon autocorrelation measurement, which is time intensive and not currently feasible with the automated process.

\hfill \\
\section{NV$^-$/NV$^0$ charge state ratio}
To determine the preferred NV charge state, we look at the ratio of NV$^-$ and NV$^0$ ZPL intensities at $\lambda$=637 and 575\,nm respectively. The NV spectra are measured under off-resonant excitation at 12\,K. NV$^-$ charge state is predominant throughout the observed excitation range (NV$^-$/NV$^0$ typically $>$\,2) for implanted NV centers in both samples A and B.

\begin{figure}[!h]
\centering
\includegraphics[width=0.35\textwidth]{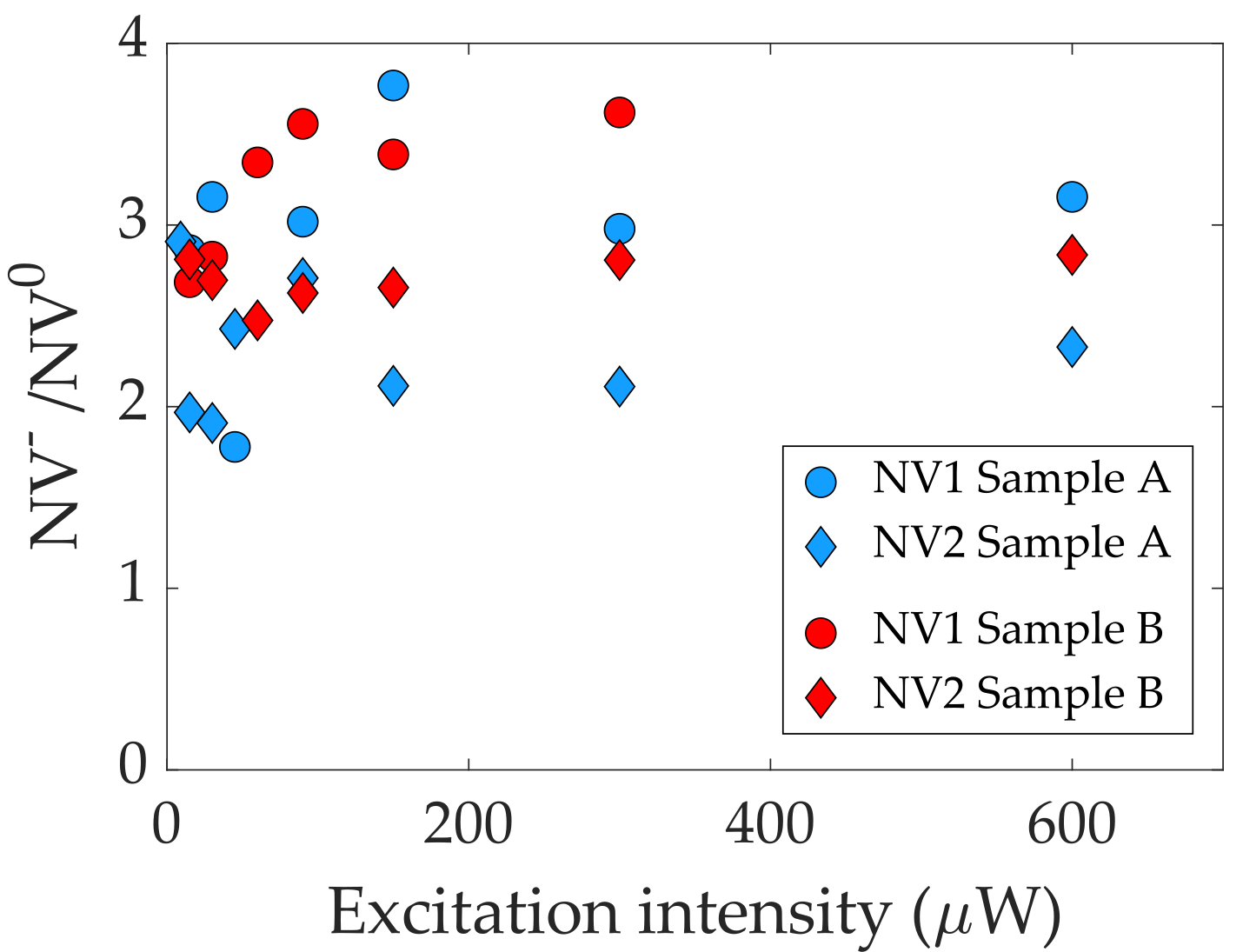}
\caption{NV charge state ratio as a function of the 532\,nm excitation intensity.}
\label{fig:charge_state_ratio}
\end{figure}

}

\bibliography{srivatsa,christian}% Produces the bibliography via BibTeX.
\end{document}